\documentclass[runningheads]{llncs}
\usepackage{url}
\usepackage{svg}
\usepackage{float}
\usepackage{ctable}
\usepackage{subfig}
\usepackage{amsmath}
\usepackage{graphicx}
\usepackage{multicol}
\usepackage{footmisc}
\usepackage{fancyhdr}
\usepackage{hyperref}
\usepackage[justification=centering]{caption}
\usepackage[ruled, noline, linesnumbered]{algorithm2e}
\fancypagestyle{specialfooter}{%
    \fancyhf{}

    \fancyfoot[CE,CO]{Accepted at \href{http://icrtc.in/}{ICRTC-2020}, to be published in \href{https://www.springer.com/series/15179}{Springer LNNS}}
}

\begin{document}

\title{Eigenvector Component Calculation Speedup over \emph{NumPy} for High-Performance Computing}
\titlerunning{Speeding up of Eigenvector Component Calculation for HPC}
%
\author{Shrey Dabhi\inst{1,2}\orcidID{0000-0002-4364-9892} \and Manojkumar Parmar\inst{1,3}\orcidID{0000-0002-1183-4399}}
\authorrunning{Shrey Dabhi et al.}
%
\institute{Robert Bosch Engineering and Business Solutions Private Limited, Bengaluru, India \\ \email{manojkumar.parmar@bosch.com} \and Department of Computer Science and Engineering, Institute of Technology, Nirma University, Ahmedabad, India \\ \email{16bit039@nirmauni.ac.in} \and HEC Paris, Jouy-en-Josas Cedex, France \\ \email{manojkumar.parmar@hec.edu}}
\maketitle              
\begin{abstract}
Applications related to artificial intelligence, machine learning, and system identification simulations essentially use eigenvectors. Calculating eigenvectors for very large matrices using conventional methods is compute-intensive and renders the applications slow. Recently, the \emph{Eigenvector-Eigenvalue Identity} formula promising significant speedup was identified. We study the algorithmic implementation of the formula against the existing state-of-the-art algorithms and their implementations to evaluate the performance gains. We provide a first of its kind systematic study of the implementation of the formula. We demonstrate further improvements using high-performance computing concepts over native \emph{NumPy} eigenvector implementation which uses LAPACK and BLAS.

\keywords{Eigenvalues \and Eigenvectors \and Eigenspace \and Symmetric matrix \and Hermitian matrix \and NumPy \and LAPACK \and BLAS \and High-performance computation \and Vectorization \and Parallelization}
\end{abstract}

\thispagestyle{specialfooter}

\section{Introduction}
In mathematics, eigenvectors are fundamental to many matrix operations. They form the base of all the dimensionality reduction operations in Artificial Intelligence and Machine Learning. Moreover, eigenvectors are of extreme importance in system simulation and calculation of many real-world phenomena. They are the key to analyzing the physics of rotating bodies, the stability of physical structures, oscillations of vibrating bodies, computational biology, economics, etc.

In linear algebra, an eigenvector is the characteristic vector of a linear transformation. It is a nonzero vector that changes at most by a scalar factor when that linear transformation is applied to it. The corresponding eigenvalue is the factor by which the eigenvector is scaled. Geometrically, an eigenvector, corresponding to a real nonzero eigenvalue, points in a direction in which it is stretched by the transformation and the eigenvalue is the factor by which it is stretched. If the eigenvalue is negative, the direction is reversed. This complexity causes the computations and thereby the applications to slow down.

In the following sections, we discuss our approach and the processes we applied to conclude. In Section 2 we explain the concept of eigenvalues and eigenvectors in brief. We also discuss the drawbacks of conventional methods and the current state-of-the-art algorithms. We then review the existing implementations of the \emph{Eigenvector-Eigenvalue Identity} formula to formulate our strategy for the systematic improvement and evaluation. In Section 3 we shed light on the actual experimental setup and present algorithms for the baseline (given by Denton et al. \cite{Denton:2019pka}\cite{denton_2020}) and the most performant variants of the implementations out of all the variants that we tested. We also present our thoughts on why exhibit Algorithm~\ref{alg:1} gives the highest amount of speed up over \emph{NumPy}'s implementation. In Section 4 we discuss the results of our experiments. In Section 5 we give the direction for future researchers to improve over our work.

\section{Related Works}

The eigenvalues of an $n \times n$ matrix $A$ are the roots of the characteristic equation:
\begin{equation}
    det(A - \lambda I) = 0\label{eq:0}
\end{equation}
where $I$ is an $n \times n$ identity matrix Eq.~\eqref{eq:0} has $n$ solutions. The polynomial left-hand side of the characteristic equation is known as the characteristic polynomial. Hermitian matrices have real eigenvalues, i.e. the roots of Eq~\eqref{eq:0} are real. A real and symmetric matrix is simply a special case of a hermitian matrix. Traditionally, eigenvectors are calculated substituting the value of $\lambda$ in the characteristic polynomial. These polynomials are $n^{th}$-order polynomials and solving them directly to calculate the eigenvectors can be very compute-intensive.

Currently, the state-of-the-art algorithm for the computation of eigenvectors of hermitian matrices is the divide and conquer algorithm proposed by Cuppen et al. \cite{Cuppen1980}. There are only 2 known implementations of this algorithm: one written in FORTRAN \cite{laug} and another one written in C language \cite{masyagin_2019}. This algorithm first converts the hermitian or real symmetric matrix into a tridiagonal matrix, and then recursively divides the matrix into 2 tridiagonal matrices until the lowest possible size is reached. Then it calculates the eigenvectors for the smallest matrices and adds rank 1 corrections while accumulating the results to compensate for the loss of information incurred during the division of the matrix. The original recursive algorithm was later improved upon by various researchers for increasing the speed \cite{Rutter:CSD-94-799}\cite{doi:10.1137/GuSC}\cite{doi:10.1137/TFD}.

A very simple formula has been recently brought to light again after it was first discovered by Jacobi in 1834 and then subsequently rediscovered by many independent research teams throughout history. The new formula is proposed by Denton et al. \cite{Denton:2019pka} and  formally named as the ``\emph{Eigenvector-Eigenvalue Identity}''. They have traced the appearance of identity \eqref{eq:1} throughout history and to prevent further disappearance and rediscovery of the identity, decided to formalize it finally in the form mentioned above. Various mathematical proofs of the identity have been covered in \cite{Denton:2019pka}, however, computational gains have not been investigated systematically. The formal definition of identity has been given below.

If $A$ is an $n \times n$ hermitian matrix with eigenvalues $\lambda _{i}(A),\dots,\lambda _{n}(A)$  and $i,j = 1,\dots, n$, then the $j^{th}$ component $v _{i,j}$ of a unit eigenvector $v _{i}$ associated to the eigenvalue $\lambda _{i}(A)$ is related to the eigenvalues $\lambda _{1}(M _{j}),\dots,\lambda _{n-1}(M _{j})$ of the minor $M _{j}$ of $A$ formed by removing the $j^{th}$ row and column by the formula
\begin{equation}
|v_{i,j} |^{2} = \frac{\prod ^{n-1}_{k=1}(\lambda _{i}(A) - \lambda _{k}(M_{j}))}{\prod ^{n}_{k=1, k\neq i}(\lambda _{i}(A) - \lambda _{k}(A))}\label{eq:1}
\end{equation}

In its current form identity \eqref{eq:1} does not provide the direction of the components of the eigenvectors, but just the magnitude. This makes it unsuitable for applications requiring directions. But as mentioned in \cite{Denton:2019pka}, it is possible to infer the directions of the components through various methods. It has been suggested by Ashok et al. \cite{Mukherjee1989} that for small matrices, the directions of eigenvectors can often be inferred by direct inspection of the eigenvector equation $A v _{i} = \lambda _{i} (A) v _{i}$. In general, we can apply Eq.~\eqref{eq:1} on multiple bases to retrieve the direction.

The basic implementation for identity formula (Eq.~\eqref{eq:1}) was provided by Denton et al. \cite{Denton:2019pka}\cite{denton_2020}. We improved upon this implementation by vectorizing, parallelizing and adding batched processing capabilities. Despite the observations of Amdahl's law, we did not achieve tangible speed-up in execution time in certain cases, due to the possible overhead of thread management.\footnote{Amdahl's Law gives the theoretical speedup in the execution time of a routine at a fixed workload that can be expected out of a system whose resources are improved. In our case, we are increasing the computational power of the system by allowing our routine to utilize multiple CPU cores in parallel.}

In this paper, we focus on implementing identity formula \eqref{eq:1} as efficiently as possible and attempt to match the accuracy, speed, and robustness of the state-of-the-art implementations. The majority of the popular machine learning and linear algebra libraries use the low-level FORTRAN implementations from BLAS packages and LAPACK under the hood, for eigendecomposition. All the machine learning libraries just provide high-level wrappers which abstract out a lot of complexities and provide sensible defaults for the required hyper-parameters to achieve the desired speed and accuracy from the eigendecomposition algorithms from LAPACK.

\section{Experiments}

The baseline implementation of identity formula Eq~\eqref{eq:1} as provided by Denton et al. \cite{denton_2020} is quite slow as it repeatedly computes eigenvalues of the same set of matrices, implying that it does not provide any advantage over other existing algorithms. The pseudo-code for the baseline implementation is provided in exhibit Algorithm~\ref{alg:0}. We used concepts of high-performance computing to improve on it. We conceived, designed and implemented 5 variations of the algorithm and the baseline version. The comparison of the performance of each version for 3 different tasks is discussed in the results section. The pseudo-code for the best performing variant is given in exhibit Algorithm~\ref{alg:1}.

Charis et al. \cite{gyurgyik_2019} provide an alternative implementation for the formula \eqref{eq:1}. They calculate only a particular eigenvector for the given matrix. A similar implementation in Python does not provide improvements over \emph{NumPy} \cite{oliphant2006guide}\cite{5725236}.\footnote{One of the reasons can be the lack of native parallel for loops in Python.} We then decided to test the speed up in the calculation of one of the components of a particular eigenvector of the given matrix.

\begin{algorithm}
\DontPrintSemicolon
    \SetKwFunction{FSub}{EigenComponentBaseline}
    \SetKwProg{Fn}{Function}{:}{\KwRet component}
    \Fn{\FSub{$matrix$, $i$, $j$}}{
        $n \gets$ matrix.shape\;
        $minor \gets$ del($matrix$, $i$)\;
        $matrixEV \gets$ EigenValues($matrix$)\;
        $minorEV \gets$ EigenValues($minor$)\;
        $numerator \gets 1.0$\;
        \For{$k = 0;\ k < n - 1;\ k \gets k + 1$}{
            $numerator \gets numerator * (matrixEV[j] - matrixEV[k])$\;
        }
        $denominator \gets 1.0$\;
        \For{$k = 0;\ k < n;\ k \gets k + 1$}{
            \If{$k \neq j$}{
                $denominator \gets denominator * (matrixEV[j] - minorEV[k])$\;
            }
        }
        $component \gets numerator \div denominator$
    }
    \caption{\label{alg:0} Baseline pseudo-algorithm}
\end{algorithm}

\begin{algorithm}
\DontPrintSemicolon
    \SetKwFunction{FSub}{EigenComponentOptimized}
    \SetKwProg{Fn}{Function}{:}{\KwRet component}
    \Fn{\FSub{$matrix$, $i$, $j$, $batchSize$}}{
        $n \gets$ matrix.shape\;
        $minor \gets$ del($matrix$, $i$)\;
        $matrixEV \gets$ EigenValues($matrix$)\;
        $eigenValue \gets matrixEV[j]$\;
        $matrixEV \gets$ del(matrixEV, $j$)\;
        $minorEV \gets$ EigenValues($minor$)\;
        $batches, nBatch \gets$ PrepareBatches($matrixEV, minorEV, batchSize$)\;
        \For{$k = 1;\ k < nBatch;\ k \gets k + 1$}{
            $tNumer[k], tDenom[k] \gets dispatch(BatchProcessor(batches[k]))$\;
        }
        $component \gets 1.0$\;
        \For{$k = 0;\ k < nBatch;\ k \gets k + 1$}{
            $component \gets component * (join(tNumer[k]) \div join(tDenom[k]))$\;
        }
    }
    \caption{\label{alg:1}Optimized pseudo-algorithm}
\end{algorithm}

The experiments to measure the speed-up were carried out on a high-performance workstation running Windows 10 operating system. It is equipped with 32 GiB of random access memory and Intel Xeon\textsuperscript{\textregistered} CPU E3-1270 v6 @ 3.8 GHz with 4 physical cores and 8 logical cores.

We use Python 3.7.6 compiled for MSC v.1916 for a 64-bit processor. The computation was run 10 times for each matrix size using each variation of the implementation. The execution time was measured using cProfile utility provided in Python's standard library. The mean values for the 10 runs have been reported.

For calculating all the components of all the eigenvectors, the baseline implementation is the slowest. For an $n \times n$ matrix, the algorithm calls the function $2n^3$ times as shown in exhibit Algorithm~\ref{alg:0}. By vectorizing the algorithm we can reduce the number of calls, therefore providing speed up.

We expected that parallel computation would reduce the execution time, but the possible overhead of thread creation and management leads to an increase in the time complexity.

The baseline and vectorized versions have limited performance for very large matrices (of the order of $150 \times 150$ and greater). One of the reasons could be the intermediate calculations leading to a datatype underflow or overflow. We batched all the calculations to combat this issue and also checked if computing all the batches in parallel would lead to any significant performance gains as well.

LAPACK's implementation of Cuppen's algorithm \cite{Cuppen1980} used by \emph{NumPy} \cite{oliphant2006guide}\cite{5725236} only returns the complete set of all the eigenvectors. Hence, we further explored if we could benefit from the calculation of single vectors or a single component of any one vector.  

\section{Results}

In Fig. 1(a) and 1(b), identity refers to the batched vectorized implementation of the Eq.~\eqref{eq:1} and identity parallelized refers to exhibit Algorithm~\ref{alg:1}. In all the figures, X-axis represents the size of the matrix for which we are calculating the eigendecomposition and the Y-axis represents the time taken in seconds.

Fig. 1(a) shows the speed up over \emph{NumPy}'s implementation for the calculation of a single component of a particular eigenvector for the given matrix. Fig. 1(b) shows the performance of our implementation against the current state-of-the-art, i.e. \emph{NumPy}, for calculating a complete eigenvector for the given matrix.

Fig. 1(c) and Fig. 1(d) show how we step-by-step implemented the concepts of high-performance computing to improve the performance of the formula. We achieved speedup with each iteration of optimization and ended up with exhibit Algorithm~\ref{alg:1} as the most optimized and performant implementation of the formula \eqref{eq:1}.

\begin{figure*}[ht]
    \centering
    \subfloat[b][For one of the components of an eigenvector]{\includegraphics[width=0.49\textwidth]{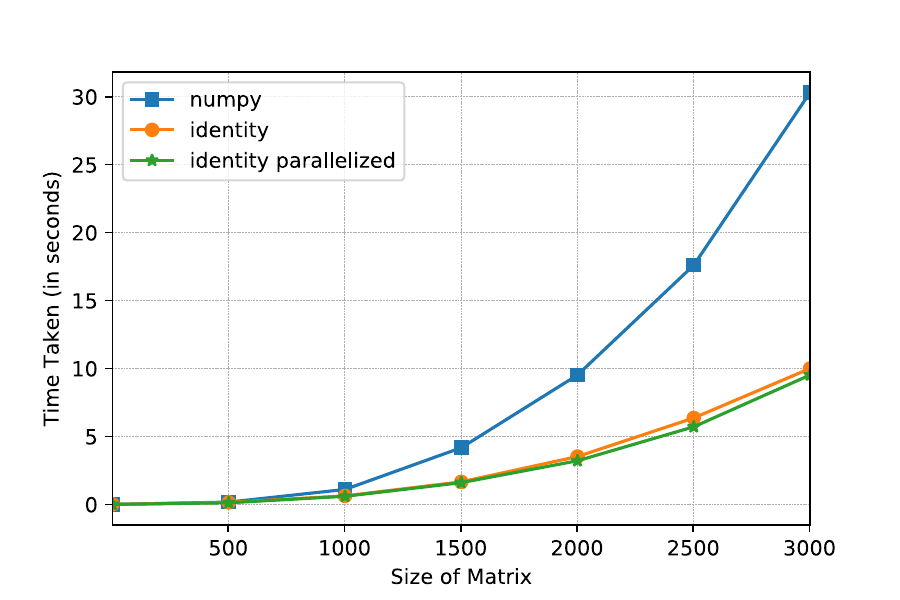}}
    \subfloat[b][For all the components of an eigenvector]{\includegraphics[width=0.49\textwidth]{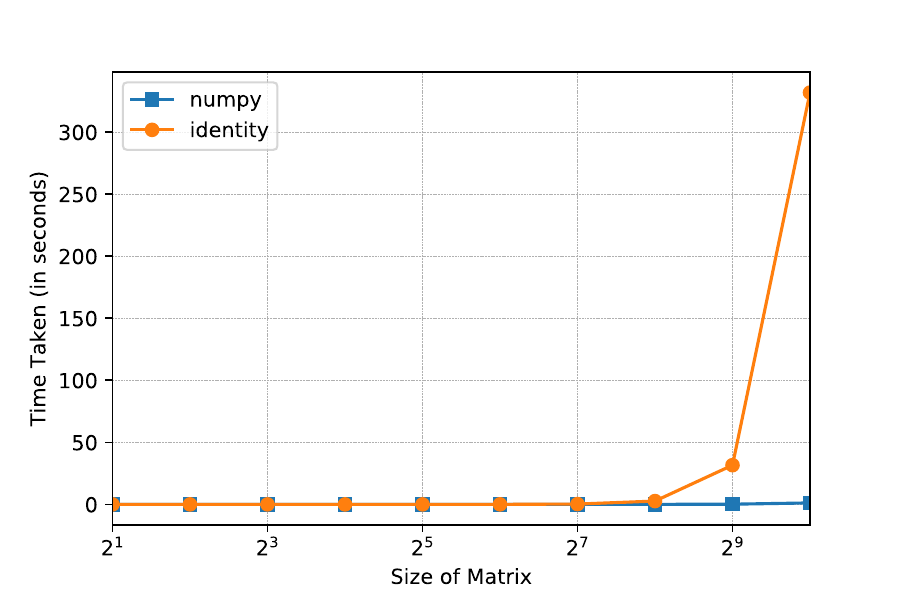}}
    \vspace{0pt}
    \subfloat[b][For all the eigenvectors]{\includegraphics[width=0.49\textwidth]{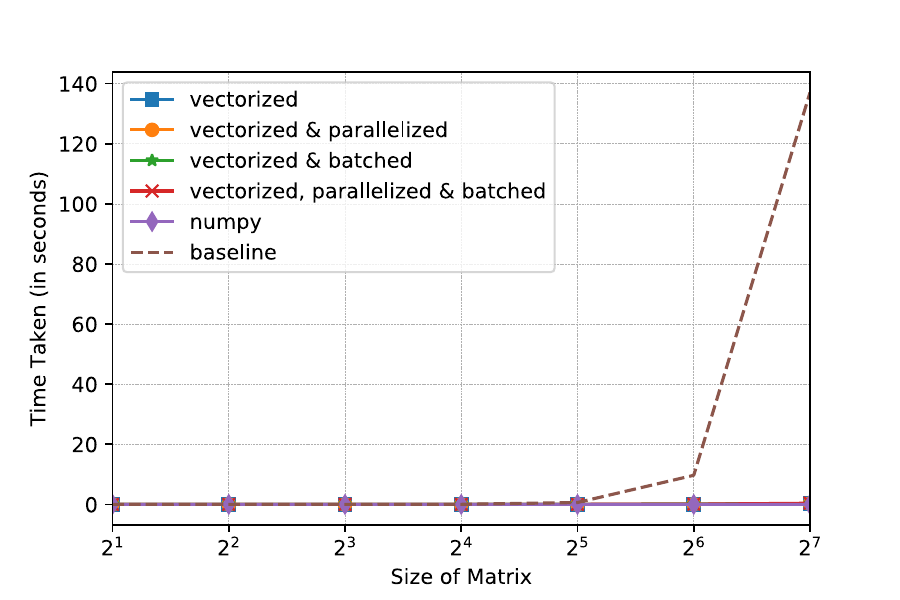}}
    \subfloat[b][For all the eigenvectors]{\includegraphics[width=0.49\textwidth]{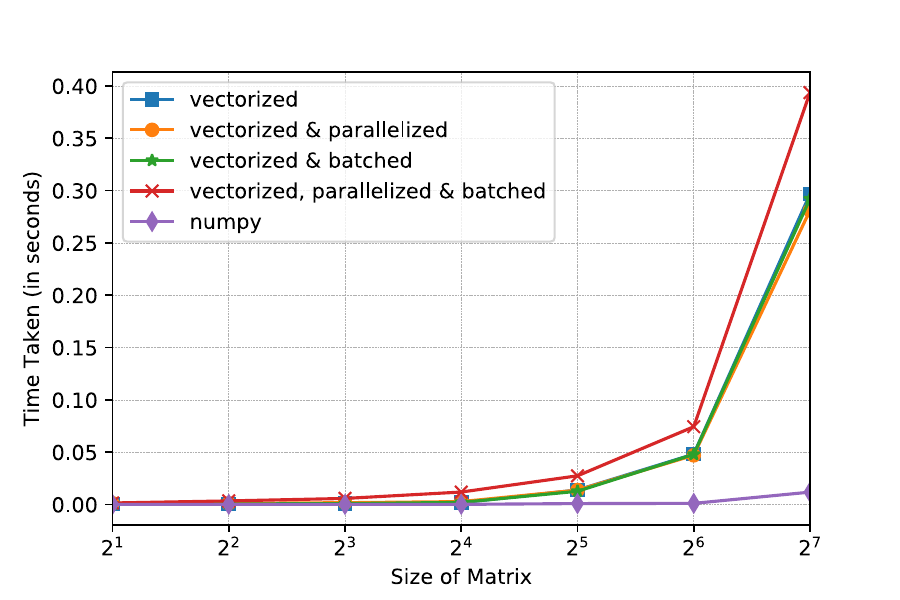}}
    \caption{Comparison of the execution time of all the variants of the algorithm and LAPACK's implementation of Cuppen's algorithm}
    \label{fig:rtimes}
\end{figure*}

From Table~\ref{tab:0} we can infer that by using Eq.~\eqref{eq:1} and exhibit Algorithm~\ref{alg:1} with an increase in the size of the matrix, we can achieve up to $4.5\times$ speed up over \emph{NumPy}.

\begin{table}[ht]
\caption{\label{tab:0} Time in seconds for calculating 1 eigenvector component}
\begin{tabular*}{\textwidth}{c @{\extracolsep{\fill}} cc}
\specialrule{.2em}{.1em}{.1em}
Size of matrix & NumPy\footnotemark & Exhibit Alg.~\ref{alg:1} \\ \specialrule{.1em}{.05em}{.05em}
2 & 0.000057 & 0.000233 \\ \hline
502 & 0.170962 & 0.119722 \\ \hline
1002 & 1.100800 & 0.595935 \\ \hline
1502 & 4.165680 & 1.603579 \\ \hline
2002 & 9.522480 & 3.212001 \\ \hline
2502 & 17.632952 & 5.707374 \\ \hline
3002 & 30.342656 & 9.522745 \\ \hline
3502 & 47.404600 & 11.886900 \\ \hline
4002 & 69.955400 & 16.775200 \\ \hline
4502 & 98.680800 & 22.943500 \\ \hline
5002 & 134.324200 & 30.547200 \\ \hline
5502 & 177.629000 & 39.741700 \\ \hline
6002 & 229.338600 & 50.682400 \\ \specialrule{.2em}{.1em}{.1em} 
\end{tabular*}
\end{table}

\footnotetext{Always computes the entire set}

The formula is more efficient than the state-of-the-art algorithms only when we need a few components of a particular eigenvector. Currently, only the applications such as web indexing, web search, signal pre-processing, etc. which only require partial eigenvectors, can benefit from the implementation of this identity.

\section{Future Work}

It might be possible to further improve the performance if the formula is implemented in a lower-level language like C or FORTRAN to reduce overheads. LAPACK has also been written in FORTRAN, and therefore it can achieve such high performance. LAPACK algorithms also benefit from clever machine level multi-threading, which is extremely limited in a higher level language like Python, despite that we are getting a glimpse of the potential improvements as seen in Fig.~\ref{fig:rtimes} (a).

\section{Summary}

The \emph{Eigenvector-Eigenvalue Identity} formula theoretically holds a promise to achieve speedup over the current state-of-the-art algorithms for the calculation of eigenvectors. We tried to realize this promise in an optimized algorithm. Our implementation can achieve up to $4.5\times$ speed up over \emph{NumPy} for applications requiring partial eigenvectors. We are of the opinion that this algorithm can further provide a speedup if implemented using lower-level languages.

\section*{Acknowledgment}

We want to thank Sri Krishnan V, Mohan B V, Palak Pareshkumar Sukharamwala and Tanya Motwani from Robert Bosch Engineering and Business Solutions Private Limited, India, for their valuable comments, contributions and continued support to the project. We are grateful to all experts for providing us with their valuable insights and informed opinions ensuring completeness of our study.
%
%
%
%
\bibliographystyle{splncs04}
\bibliography{ref}

\end{document}